\documentstyle[10pt]{article}

\title{
Comment on ``Equal-time temperature correlators of the one-dimensional 
Heisenberg XY chain'', preprint solv-int/9710028}

\author{
K.~N.~Ilinski\thanks{E-mail: KNI@TH.PH.BHAM.AC.UK},
A.~S.~Stepanenko\thanks{E-mail: ASS@TH.PH.BHAM.AC.UK}
\\
{\small\it School of Physics and Astronomy, University of
Birmingham,}
\\
{\small\it Edgbaston, Birmingham B15 2TT, United Kingdom.}}
\date{ }

\begin{document}

\maketitle

\begin{abstract}
In the comment we give references to our papers where the problem
was solved for more general case of time-dependent finite temperature
correlators. 
\end{abstract}

\thispagestyle{empty}

In the recent preprint solv-int/9710028 \cite{IKK} by Izergin, 
Kapitonov and  Kitanine representations in terms of 
determinants of 
$M\times M$ matrices are obtained for equal time temperature 
correlators of the anisotropic XY chain. We would like to note
that the problem is a special case of the more general problem 
of the calculation 
of general time-dependent correlators for the XY chain. The last 
problem was  solved in our paper \cite{IIKMS} and 
partially reported in \cite{IK}. 

In paper \cite{IIKMS} we obtained exact results on the 
linear response of cyclic molecular aggregates. Using the fact 
that in the dipole-dipole approximation the Hamiltonian of 
cyclic molecular aggregates reduces to the anisotropic XY chain,
we mapped the problem to the one of the calculation of general 
time-dependent correlators. In \cite{IIKMS} we described 
the coherent state technique of calculation of the correlators
which is exactly the same as the one used (two years later) in 
\cite{IKK} without any reference. In paper \cite{IIKMS}
explicit expressions for general time-dependent two-point 
correlation functions in the form of pfaffians of $(2M)\times (2M)$
matrices (which can be  easily reduced to determinants of 
$M\times M$ matrices) were obtained. It was shown that in the 
thermodynamical limit and in absence of the anisotropy our 
expressions are in agreement with the results obtained earlier 
for XX0
chain. For  small numbers of sites the results have been checked 
symbolically and numerically.

In conclusion, the authors of Ref.~\cite{IKK} claim to reobtain 
results known from our papers Ref.~\cite{IIKMS}. 
However, it appears strange that the only mention of 
our results was found in the last paragraph of the last page of 
the paper without any comments or comparisons. The last 
sentence of preprint solv-int/9710028 ``In our next paper we hope to 
give clear answers for the time-dependent correlators" must imply
 a correction to the list of references.
This is even more questionable since the authors of the 
aforementioned preprint were aware of the existence of 
Ref.~\cite{IIKMS}.


\begin{thebibliography}{9}

\bibitem{IKK} A.G. Izergin, V.S. Kapitonov, N.A. Kitanine,
``Equal-time temperature correlators of the one-dimensional 
Heisenberg XY chain'', preprint solv-int/9710028

\bibitem{IIKMS} A.~V.~Ilinskaia, K.~N.~Ilinski,
G.~V.~Kalinin, V.~V.~Melezhik, A.~S.~Stepanenko,
``Exact results on linear response of cyclic molecular aggregates'',
preprint cond-mat/9509040.

\bibitem{IK} K.~N.~Ilinski, G.~V.~Kalinin, 
``Cyclic XY model and exotic statistics in one dimension'',
Phys.Rev. E54 (1996), R 1017--R 1020.

\end{thebibliography}
\end{document}